\documentclass{INTERSPEECH2023}

\usepackage{amsmath,graphicx,tabularx}
\usepackage{booktabs}
\usepackage{xcolor}
\usepackage{caption}
\usepackage{subcaption}
\usepackage{comment}
\usepackage{svg}
\usepackage{amsmath}
\usepackage{adjustbox}
\usepackage{enumitem}

\usepackage{caption}
\usepackage[activate]{microtype}
\newcommand{\blue}[1]{\textcolor{black}{#1}}

\renewcommand{\textcolor}[2]{#2}

\title{Voice Conversion for Lombard Speaking Style with Implicit and Explicit Acoustic Feature Conditioning}

\name{Dominika Woszczyk$^1$$^*$\thanks{$^*$Work done while at Amazon Alexa TTS Research.}, Manuel Sam Ribeiro$^2$, Thomas Merritt$^{2*}$, Daniel Korzekwa$^{2*}$}

\address{$^1$Department of Computing, Imperial College London, London, UK\\
$^2$Amazon Alexa, TTS Research, UK}

\email{$^1$d.woszczyk19@imperial.ac.uk}

\interspeechcameraready
\begin{document}

\maketitle

\begin{abstract}
Text-to-Speech (TTS) systems in Lombard speaking style can improve the overall intelligibility of speech, useful for hearing loss and noisy conditions. However, training those models requires a large amount of data and the Lombard effect is challenging to record due to speaker and noise variability and tiring recording conditions. Voice conversion (VC) has been shown to be a useful augmentation technique to train TTS systems in the absence of recorded data from the target speaker in the target speaking style. In this paper, we are concerned with Lombard speaking style transfer. Our goal is to convert speaker identity while preserving the acoustic attributes that define the Lombard speaking style. We compare voice conversion models with implicit and explicit acoustic feature conditioning. We observe that our proposed implicit conditioning strategy achieves an intelligibility gain comparable to the model conditioned on explicit acoustic features, while also preserving speaker similarity.



\end{abstract}

\section{Introduction}
\label{sec:intro}

Traditional approaches to improving speech intelligibility for individuals with hearing aids primarily focus on signal processing and amplification at the receiving end. However, in this paper, we propose an alternative approach that seeks to enhance intelligibility by addressing the source of speech generation rather than relying solely on signal processing and amplification. Specifically, we investigate the use of Lombard speech in text-to-speech (TTS) systems to mimic the natural adjustments made by speakers in noisy environments. We argue Lombard-style TTS could be an effective way to improve the intelligibility of speech, easier to receive with hearing loss and also more generally speech in noisy environments.

In fact, the Lombard speaking style has already been applied to Text-to-Speech (TTS) systems~\cite{bollepalli2018speaking,hu2021whispered,paul2020enhancing} and has demonstrated a positive impact on the intelligibility of synthesized voices in noise~\cite{paul2020enhancing}. Nevertheless, current TTS models require data for the target speaker in the target speaking style, and Lombard speech is difficult to record. The recording conditions are tiring due to the noise, and the Lombard effect patterns and intensity vary from one speaker to another. 

In this paper, we present a data augmentation technique for Lombard text-to-speech systems. Voice conversion has been shown to be useful as a data augmentation technique for TTS, style conversion and source style transfer~\cite{ribeiro2022cross,huybrechts2021low,terashima2022cross} However, most voice conversion models focus on the speaker identity and ignore additional characteristics that are transferred with it (emphasis, emotion, intonation). Voice conversion for speaking styles is challenging as it often loses the source style. We focus on the inherent challenges of performing voice conversion for Lombard speech, in the optic to train a TTS model on it and address the problem of voice conversion preserving source Lombard speaking style using implicit and explicit acoustic feature conditioning.  Our goal is to convert speaker identity while retaining the source speakers' Lombard speaking style. By preserving the Lombard features of the original speaker in voice conversion, we can generate synthetic datasets for a target voice and ensure that the resulting audio signal sounds more intelligible and natural to the listener. Past works perform source style transfer by explicitly modeling key characteristics of the source style to preserve them ~\cite{li2020normal,lopez2017speaking, huang2010lombard,raitio2011analysis,raitio2022vocal}. However, this approach requires manual analysis and under-utilize the feature extraction capabilities of today's deep neural networks. More recently, models that implicitly learn the desired target style speech attributes were explored, either via adversarial loss~\cite{seshadri2019augmented,schnell2021emocat} or disentanglement~\cite{chan2022speechsplit2,bian2019multi} for both VC and TTS.  Style reconstruction loss is another technique that also has been applied to enforce expressive speech synthesis on TTS~\cite{liu2021expressive}. In the domain of Lombard-style voice conversion, implicit feature modeling has yet to be explored.

In our work, we focus on the \blue{problem} of intelligibility-preserving voice conversion for Lombard speaking style transfer and we propose to model the Lombard prosody implicitly using a style reconstruction loss to overcome these issues and compare it to explicit modeling. We compare implicit and explicit conditioning on a many-to-many voice conversion model. Our results show that a model with a style reconstruction loss achieves an intelligibility increase comparable to a model with explicit conditioning, whilst better preserving speaker identity.

\section{Analysis of the Lombard Speaking Style}
\label{sec:study}

\begin{figure*}[]
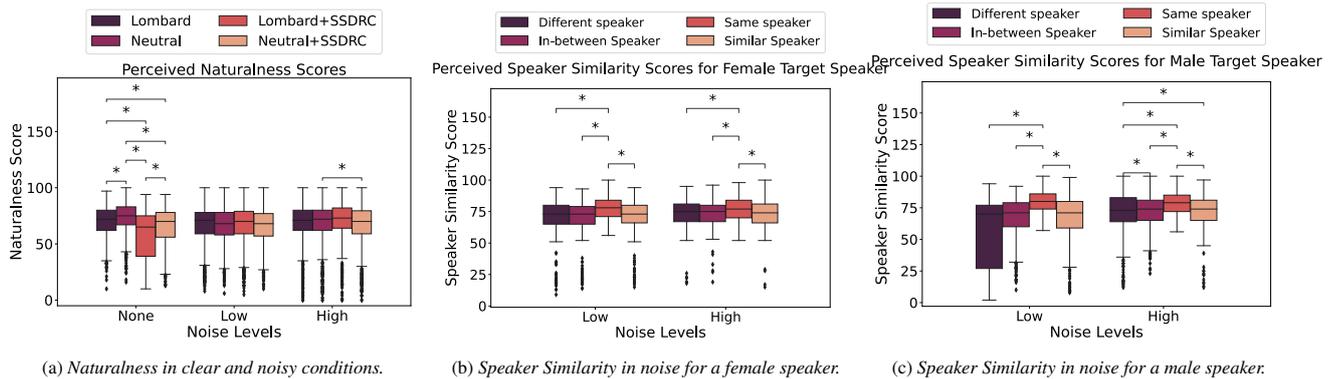

     \centering
      \begin{subfigure}[b]{0.33\textwidth}
      \centering
     \includesvg[width=6cm]{Figures/pil_nat_filt_sig.svg}
     \caption{Naturalness in clear and noisy conditions.}
      \end{subfigure}
      \hfill
     \begin{subfigure}[b]{0.33\textwidth}
         \centering
        \includesvg[width=6cm]{Figures/pil_ss_s27_filt_sig.svg}
        \caption{Speaker Similarity in noise for a female speaker.}
     \end{subfigure}
     \hfill
     \begin{subfigure}[b]{0.33\textwidth}
        \centering
         \includesvg[width=6cm]{Figures/pil_ss_s43_filt_sig.svg}
        \caption{Speaker Similarity in noise for a male speaker.}
     \end{subfigure}
        \caption{Boxplots of Naturalness and Speaker Similarity in both low (SNR-1) and high (SNR-3) noise levels from the Pilot study. The symbol * indicates significance for $p\leq0.05$.}
        \label{fig:pil_similarity}
\end{figure*}


\subsection{Background}
Signal processing approaches that are noise independent such as spectral shaping (SS) and dynamic range compression (DRC) have shown to be helpful in improving speech intelligibility~\cite{rennies2020intelligibility}. SS distributes the energy in the frequency domain, sharpens the formants and reduces the spectral tilt while DRC amplifies quiet sounds and reduces loud sounds. While spectral shaping and dynamic range compression (SSDRC) can efficiently improve intelligibility without additional data collection, they do impact the naturalness of the speech samples. Lombard speech has been shown to improve the intelligibility over natural speech~\cite{junqua1993lombard} and has been studied in the context of improving intelligibility of speech disorders and hearing aids~\cite{hansen2020speech}, or build robust automatic speech recognition systems~\cite{junqua1993lombard}. The Lombard effect has a high variability as it not only comes from the surrounding noise but also relies on the feedback received from the listener, and the speaker itself. However, general trends have been observed. Speakers tends to increase the duration of vowels and decrease their spectral tilt; increase fundamental frequency (f0); and the spectral energy of vowels is moved from high and low frequencies to mid-level frequencies. In the context of TTS systems, it is important to achieve pleasant voices for users. The impact of the Lombard effect on naturalness has been observed in recent work~\cite{hu2021whispered,vojtech2019effects}, but there has been no study that made it its focus.
Another relevant dimension to style conversion in speech synthesis is speaker similarity and voice identity in noise.
Humans are very good at detecting many cues from a person's voice and can distinguish similar speakers given similar conditions~\cite{latinus2011human}. Unfortunately studies have shown that they are not reliable when it comes to grouping speaker identity when confronted with speech samples that include different emotions, whisper, laughs or other languages~\cite{lavan2019breaking}. Different listeners also have different levels of perception and can be further affected by noise~\cite{song2011perception}.

In this section we investigate the impact of the Lombard effect in perceived intelligibility and naturalness in noise, and compare it against SSDRC. Additionally, we study the ability of listeners to perceive the speaker identity and similarity in noise. To this end, we design several pilot studies and describe them in the following sections.  


\subsection{Evaluation Setup}
\label{sec:pil_eval}

We perform our evaluation on the Audio-Visual Lombard Grid dataset~\cite{alghamdi2018corpus} which consists of 54 speakers (24 Male, 30 Female). Each speaker has a total of 100 recordings of randomly generated sentences in English (50 Lombard and 50 neutral). We evaluate the Objective Intelligibility (OI) of our systems by computing the speech intelligibility in bits (SIIB) score~\cite{van2017instrumental}, an objective metric for intelligibility in noise, given speech-shaped noise at two speech-to-noise ratio (SNR): SNR -1 and SNR -3.
We also run subjective evaluations for Perceived Intelligibility, Naturalness and Speaker Similarity with MUSHRA-like (MUltiple Stimuli with Hidden Reference and Anchor)~\cite{series2014method} listening tests.
To measure Speaker Similarity, we ask 50 listeners to compare and rate the similarity of the systems samples to a reference sample of the target speaker. For Intelligibility and Naturalness, we do not include a reference sample but ask listeners to rate how intelligible or natural the samples sound on a scale from 0 to 100 (100 =``Very Intelligible/Natural''). For each listening test, we include 10 samples from 10 speakers, balancing for gender, for a total of 100 samples.

\subsection{Results}
\label{sec:pil_intelli}
\begin{table}[!ht]
\centering
\tiny
\resizebox{\columnwidth}{!}{ 
\begin{tabular}{lcccc}
\toprule
 & \multicolumn{2}{c}{\textbf{SNR -1}} & \multicolumn{2}{c}{\textbf{SNR -3}} \\
 \cmidrule{2-3}
  \cmidrule{4-5}
  \textbf{Systems}  & OI & SI &  OI & SI  \\ 
 \midrule
Neutral &  136.30 (.99)&57.26 (1.68)*   &  69.55 (.67)&26.17 (2.10)†   \\
Lombard &  154.24 (.96)&65.66 (1.37) &  85.22 (.73) &26.03 (2.05)*†  \\
Neutral + SSDRC & 243.76 (1.14)&57.28 (1.69)*  & 148.91 (.86)&25.48 (2.02)*\\
Lombard + SSDRC &  \textbf{251.78 (1.21)}&\textbf{66.36 (1.31)}  & \textbf{150.32 (.82)} &  \textbf{42.61 (2.33)} \\
\bottomrule
\end{tabular}%
}
\caption{Objective (OI) and subjective (SI) Intelligibility in noise at SNR -1 (dB) and SNR -3 (dB) for the pilot study, measured with SIIB and MUSHRA-like test. The higher the value the better. The best system is highlighted in bold. All pairs of systems are significantly different for $p\leq 0.005$, except the row-wise pairs indicated by * and †. }
\label{tab:siib}
\end{table}

We compare neutral and Lombard speech samples with and without SSDRC. First, we aim to investigate the \textbf{Perceived Intelligibility} of speech in noise. Results are presented in Table~\ref{tab:siib}. We observe that the Lombard effect is rated more intelligible by human listeners at low noise level (SNR -1), and by the SIIB score than \textit{Neutral} for both SNR -1 and SNR -3. On the other hand, while the SIIB score seems to favour the SSDRC method, \textit{Lombard} is still ranked higher than \textit{Neutral+SSDRC} by listeners at low and high noise level. Nevertheless, the combination of \textit{Lombard+SSDRC} performs the best in terms of both subjective and objective scores for all noise levels.

We evaluate the impact of Lombard speech on \textbf{Naturalness} in a similar fashion to the Perceived Intelligibility. Results are shown on Figure~\ref{fig:pil_similarity}.a. We notice that while the Lombard effect is rated as less natural, it is still better ranked than \textit{SSDRC} approaches. However, this effect is less apparent for high noise level. \blue{These results support the hypothesis that, in noisy conditions, the importance of naturalness is lost in exchange for intelligibility~\cite{hu2021whispered,vojtech2019effects}. This indicates that while the Lombard effect might affect the naturalness, it is negligible in noise and remains advantageous for its intelligibility gains.}

Finally, we aim to understand the importance of preserving the target speaker identity during style conversion in noisy conditions, by investigating how listeners perceive \textbf{Speaker Similarity} in noise. To this aim, we pick samples from one speaker and compare them to samples from three others: a speaker with a similar voice (\textit{Same Speaker}), one with a distinctly different voice (\textit{Different Speaker}), and speaker that lies in-between (\textit{In-between Speaker}). Results for both genders are presented in Figures \ref{fig:pil_similarity}.b and \ref{fig:pil_similarity}.c. We observe that even in noise, the order of similarity between speakers remains the same. Unlike the \textbf{Naturalness} study, these results indicate that for style conversion, the speaker identity remains important to preserve.

\section{\textcolor{blue}{Lombard Style Transfer}}
\label{sec:methodology}

\begin{figure}[]
\centering
\includegraphics[width=5.5cm]{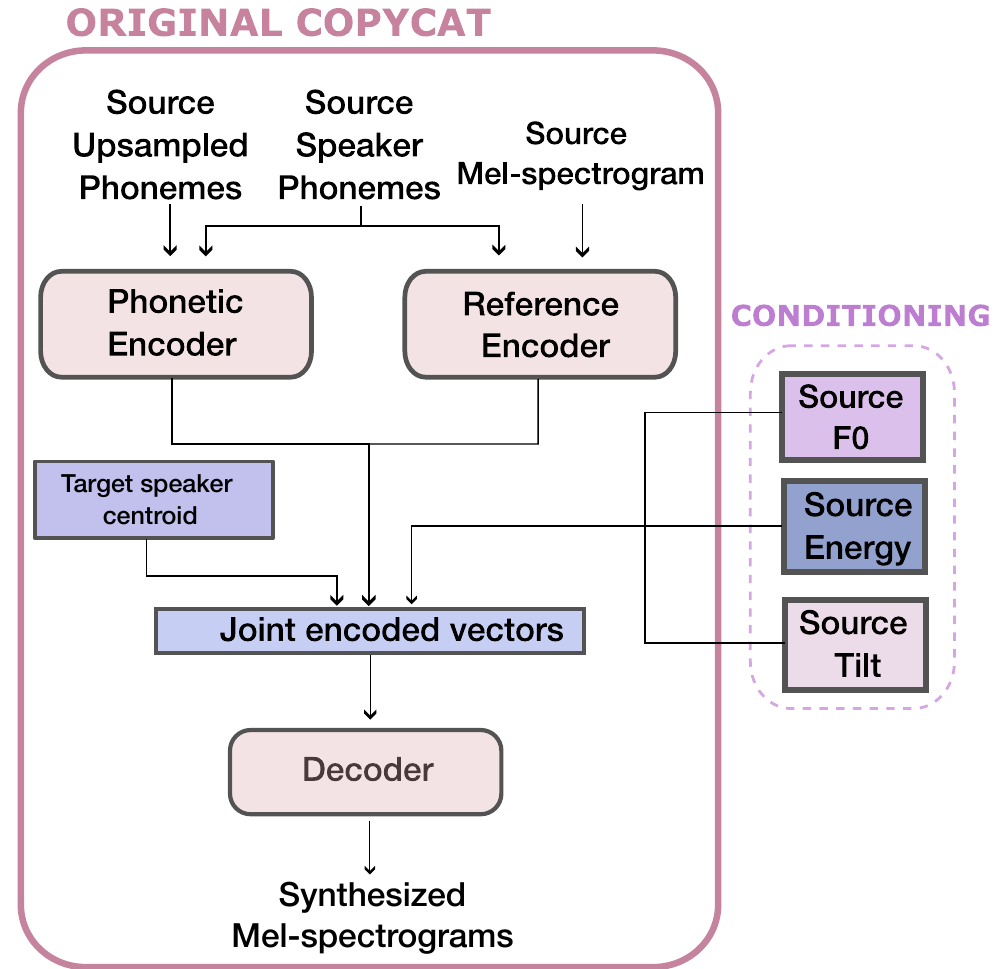}
\caption{Overall architecture of the voice conversion model, a modified version of Copycat~\cite{ribeiro2022cross, huybrechts2021low} with source features conditioning.}
\label{fig:cc_cond}
\end{figure}

\subsection{Voice Conversion Model}

Our voice conversion model is based on CopyCat~\cite{karlapati2020copycat}, a non-parallel many-to-many prosody transfer model. \blue{As shown on Figure~\ref{fig:cc_cond}}, the model has a reference encoder that takes mel-spectrograms and speaker embeddings extracted from a speaker verification model. The speaker embeddings are also passed together with phonemes upsampled at the frame level. The resulting encoded vectors are joined and given as input to a decoder that outputs decoded mel-spectrogram ready to be vocoded. Our baseline and conditioned models are the Copycat models \blue{as described in~\cite{ribeiro2022cross, huybrechts2021low}}. We use a Kullback–Leibler (KL)-divergence loss ($L_{KL}$) for the Variational Autoencoder (VAE) component in the reference encoder and an L1 loss on the source and reconstructed mel-spectrograms ($L_{rec}$). 

\subsection{Voice Conversion with Explicit Conditioning} 
\label{sec:vc_explicit}
We implement explicit acoustic conditioning for Lombard speaking style transfer, shown in Figure~\ref{fig:cc_cond}. We consider \textit{f0}, \textit{mgc0} (spectral energy) and \textit{mgc1} (spectral tilt). We extract features at frame-level using WORLD vocoder ~\cite{morise2016world} directly from the source speech waveform and feed them to the decoder, to condition the model to follow the source distributions. The target speaker embedding is the centroid of all embeddings from that speaker’s training data. We join them to the encoded feature vectors and feed it to the decoder.

\subsection{Voice Conversion with Style Reconstruction Loss}
\begin{figure}[]
\centering
\includegraphics[width=8cm]{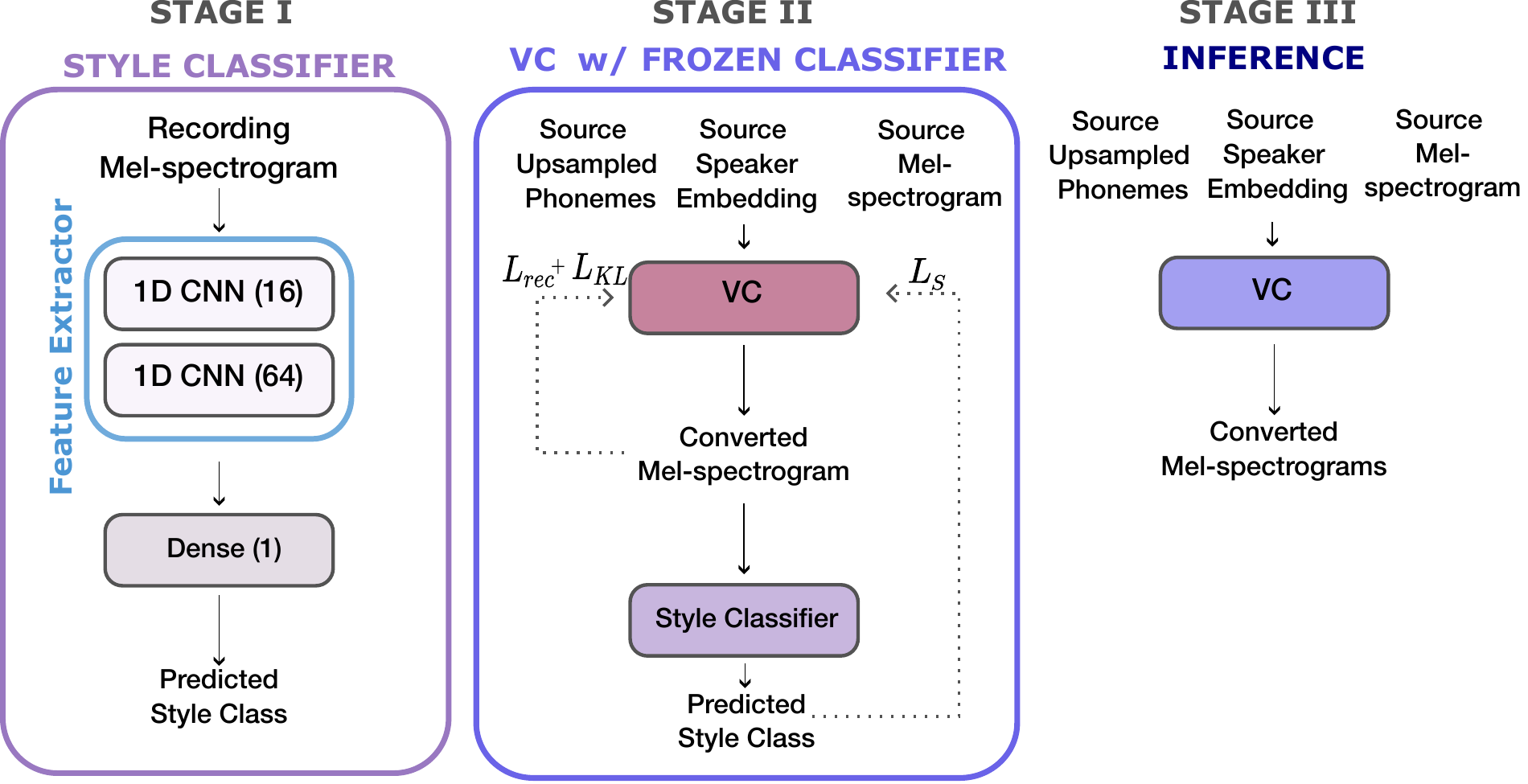}
\caption{Architecture and training schedule of the Voice Conversion (VC) Model with style classifier. Stage I is classifier training. Stage II is training VC with classifier as frozen style discriminator, and Stage III is inference with trained VC model and the style classifier is dropped. }
\label{fig:cc_clf}
\end{figure}

To enforce implicit feature learning, we add an auxiliary loss to our VC model, as shown in Figure~\ref{fig:cc_clf}. This loss forces the model to generate mel-spectrograms that preserve the source style, given the features that the style classifier learned representations during the pre-training. The training is split into two stages. First, a style classifier is trained on the Lombard/neutral recordings. Then, the weights of the classifier are frozen and the model is used to predict the class of synthesized mel-spectrograms. We use binary cross-entropy loss, which we call style reconstruction loss ($L_s$), in addition to $L_{rec}$ and $L_{KL}$. During inference, the style classifier is dropped. Finally, we evaluate a fusion model of style classifier and explicit conditioning by adding the source features from Section~\ref{sec:vc_explicit} to the model with style reconstruction loss.

\subsection{Evaluation Setup \& Results}
\label{sec:res}
\color{black}
%
In our experiments we aim to answer the following questions: 1) ``Can we preserve the key properties of the Lombard effect during voice conversion''; and 2) ``Can we preserve speaker identity''.
To evaluate our systems, we follow the same dataset and metrics setup as described in  Section~\ref{sec:pil_eval}. We train our model on 52 speakers and pick two target speakers, one female (s27) and one male (s43) speaker with high SIIB score for their neutral recordings. We train our VC model with and w/o conditioning on a many-to-many task for 100k steps. The classifier is trained on Oracle mel-spectrograms of the Lombard Grid dataset for 5k steps. We train the VC model together with the classifier with frozen weights for an additional 100k. We use a batch size of 16 and a learning rate of 0.0001. 

\begin{figure}[!ht]
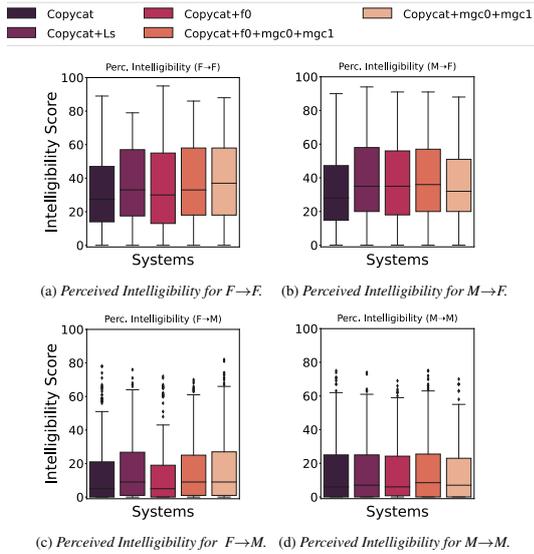

\centering
\begin{adjustbox}{minipage=\linewidth,scale=0.97}
 \begin{subfigure}[b]{1.4\columnwidth}
    \hspace{1em}
    \adjustbox{trim=4.0cm 2.8cm 0cm 0.22cm, clip}{%
    \includesvg[width=\columnwidth]{Figures/s27_cc_ss_filt_legend.svg}
    }
     \vspace{-0.25em}%
     \end{subfigure}
     \end{adjustbox}
     \centering
     \begin{adjustbox}{minipage=\linewidth,scale=0.8}
     \centering
     \begin{subfigure}[b]{0.49\columnwidth}
         \centering
        \includesvg[width=0.9\columnwidth]{Figures/s27_cc_int_f_high_clean.svg}
        \caption{Perceived Intelligibility for \textit{F$\rightarrow$F.}}
     \end{subfigure}
      \begin{subfigure}[b]{0.49\columnwidth}
         \centering
        \includesvg[width=0.83\columnwidth]{Figures/s27_cc_int_m_high_clean.svg}
        \caption{Perceived Intelligibility for \textit{M$\rightarrow$F.}}
     \end{subfigure}
     
     \begin{subfigure}[b]{0.49\columnwidth}
        \centering
        \includesvg[width=0.9\columnwidth]{Figures/s43_cc_int_f_high_clean.svg}
        \caption{Perceived Intelligibility for \textit{ F$\rightarrow$M.}}
     \end{subfigure}
     \begin{subfigure}[b]{0.49\columnwidth}
        \centering
         \includesvg[width=0.83\columnwidth]{Figures/s43_cc_int_m_high_clean.svg}
        \caption{Perceived Intelligibility for \textit{M$\rightarrow$M.}}
     \end{subfigure}
     \end{adjustbox}
        \caption{Cross-gender Perceived Intelligibility in high (SNR-3) noise settings of a VC model with different explicit conditioning and with style reconstruction loss ($L_S$). Higher is better. 
        } 
        \label{fig:cross_cc}
\end{figure}

\begin{figure}[!ht]
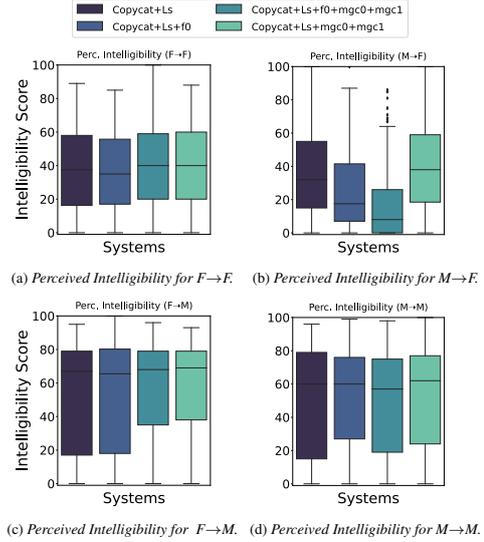

\centering
\begin{adjustbox}{minipage=\linewidth,scale=0.97}
 \begin{subfigure}[b]{\columnwidth}
  \centering
    \adjustbox{trim=3.4cm 2.5cm 0cm 0.22cm, clip}{%
    \includesvg[width=\columnwidth]{Figures/s43_clf_int_f_filt_legend.svg}
    }
    \vspace{-0.25em}%
     \end{subfigure}
     \end{adjustbox}
     \centering
     \begin{adjustbox}{minipage=\linewidth,scale=0.8}
     \centering
     \begin{subfigure}[b]{0.49\columnwidth}
         \centering
        \includesvg[width=0.9\columnwidth]{Figures/s27_clf_int_f_high_clean.svg}
        \caption{Perceived Intelligibility for \textit{F$\rightarrow$F.}}
     \end{subfigure}
      \begin{subfigure}[b]{0.49\columnwidth}
         \centering
        \includesvg[width=0.82\columnwidth]{Figures/s27_clf_int_m_high_clean.svg}
        \caption{Perceived Intelligibility for \textit{M$\rightarrow$F.}}
     \end{subfigure}
     
     \begin{subfigure}[b]{0.49\columnwidth}
        \centering
        \includesvg[width=0.9\columnwidth]{Figures/s43_clf_int_f_high_clean.svg}
        \caption{Perceived Intelligibility for \textit{ F$\rightarrow$M.}}
     \end{subfigure}
     \begin{subfigure}[b]{0.49\columnwidth}
        \centering
         \includesvg[width=0.82\columnwidth]{Figures/s43_clf_int_m_high_clean.svg}
        \caption{Perceived Intelligibility for \textit{M$\rightarrow$M.}}
     \end{subfigure}
     \end{adjustbox}
        \caption{Cross-gender Perceived Intelligibility in high (SNR-3) noise settings of the fusion of a VC model with style reconstruction loss ($L_S$) and explicit conditioning. Higher is better. 
        }
        \label{fig:cross_clf}
\end{figure}

\begin{figure}[!ht]
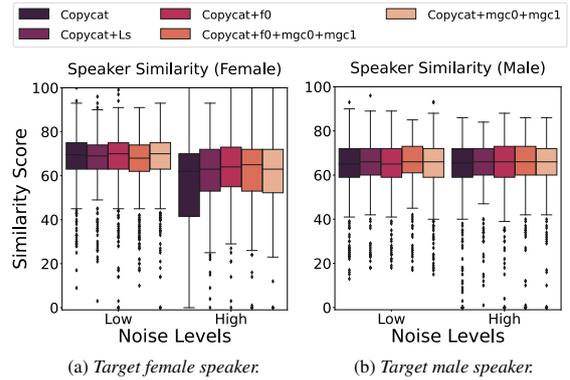

\begin{adjustbox}{minipage=\linewidth,scale=1}
\begin{subfigure}[b]{1.4\columnwidth}
    \adjustbox{trim=3.9cm 2.8cm 0cm 0.00cm, clip}{%
    \includesvg[width=\columnwidth]{Figures/s27_cc_ss_filt_legend.svg}
    }
    \vspace{-0.25em}%
     \end{subfigure}
    \end{adjustbox}
    \centering
    \begin{adjustbox}{minipage=\linewidth,scale=1.0}
     \begin{subfigure}[b]{0.5\columnwidth}
         \centering
        \includesvg[width=\columnwidth]{Figures/s27_ss_filt_legend_clean.svg}
        \caption{Target female speaker.}
     \end{subfigure}
    \hspace{-1em}%
     \begin{subfigure}[b]{0.48\columnwidth}
        \centering
        \includesvg[width=\columnwidth]{Figures/s43_ss_filt_clean.svg}
        \caption{Target male speaker.}
        \end{subfigure}
    \end{adjustbox}
     \caption{Perceived Speaker Similarity in low (SNR-1) and high (SNR-3) noise settings of a VC model with explicit conditioning and with style reconstruction Loss ($L_S$). Higher is better. 
     }
    \label{fig:spk_similarity}
\end{figure}

We summarize the results for the objective intelligibility measure for explicit and implicit conditioning systems as well as their fusion in Table~\ref{tab:intelligililbity}. Figures~\ref{fig:cross_cc} and~\ref{fig:spk_similarity} present the results of the subjective intelligibility in high noise and speaker similarity evaluation. We perform an independent t-test with a Holm-Bonferroni correction for all evaluations. We also evaluate them in low noise settings, and we observed that adding source features or enforcing the style reconstruction helped to improve the intelligibility of the samples for all systems, as all models were significantly better than the base \textit{VC} model. However, in high noise settings the improvements are less apparent, as shown in Figure~\ref{fig:cross_cc}. We observe differences given the source speaker's gender. For female source speakers, the best system is \textit{mgc0+mgc1}, and \textit{f0} seems to be harmful to intelligibility. On the other hand, for male source speakers, both \textit{VC + Ls} and \textit{f0+mgc0+mgc1} are significantly better than the baseline. In general, \textit{mgc0} and \textit{mgc1} features perform the best. However, the $f0$ seems to be beneficial for male source speakers. Nevertheless, looking at Figure~\ref{fig:cross_cc} and Table~\ref{tab:intelligililbity}, the model with implicit conditioning performs similarly to \textit{Ls+f0+mgc0+mgc1}. Additionally, looking at speaker similarity in Figure~\ref{fig:spk_similarity}, we see that the baseline \textit{VC} transfers more of the target speaker identity but that \textit{Ls} perform similarly to other models. One can see on Figure~\ref{fig:cross_clf} that combining only some of the few explicit features with the $L_s$ loss seems to either not add much or be detrimental to the intelligibility. On the other hand, the fusion of all of them improves the performance with the best results achieved with \textit{Ls+mgc0+mgc1}.

\newcommand*{\MyIndent}{\hspace*{0.3cm}}%

\begin{table}[!ht]
\small
\centering
\resizebox{\columnwidth}{!}{ 
\begin{tabular}{lcc|cc}
\toprule
& \multicolumn{2}{c}{\textbf{Female target}} & \multicolumn{2}{c}{\textbf{Male target}} 
\\
\cmidrule(lr){2-3}
\cmidrule(lr){4-5}
    \textbf{Systems}                 & \multicolumn{1}{c}{\textbf{SNR-1}} & \multicolumn{1}{c}{\textbf{SNR-3}}  & \multicolumn{1}{c}{\textbf{SNR-1}} & \multicolumn{1}{c}{\textbf{SNR-3}} \\

\midrule
              Source Lombard Recordings & 154.25 (.96) &85.22 (.72)& 134.53 (5.63) & 73.18 (4.83)\\
\hline & \\[-1.5ex]
VC &  83.92 (3.56)*&46.29 (2.74)*& 102.58(6.43)& 38.78 (3.77)* \\
        \MyIndent +f0 & 84.21 (3.56)*&46.61 (3.11)*& 98.46 (5.92) &38.13 (3.10)*\\
         \MyIndent+mgc0+mgc1 & 100.38 (1.80) &55.42 (1.47) &117.15 (5.7)*& 46.91 (3.00) \\
         \MyIndent+f0+mgc0+mgc1 &   93.85 (3.57) & 53.84 (2.69) &120.73 (5.87)& 49.37 (3.24) \\
\hline\\[-1.5ex]

VC + L$_{S}$ & 86.79 (3.57) &51.44 (2.8) &107.26 (4.88)  &44.14 (3.27)   \\
\MyIndent +f0 &  106.19 (4.27)& 55.23 (2.29)&\textbf{118.85 (7.64)}*&\textbf{51.89 (3.05)}\\
\MyIndent+mgc0+mgc1 & 104.37 (2.40)& 58.24 (1.97)& 87.48 (3.48) &48.83 (3.28)† \\
\MyIndent+f0+mgc0+mgc1 & \textbf{115.13 (3.19)}& \textbf{60.65 (1.90)}& 110.61 (6.95)&49.00 (3.72)†  \\
\bottomrule
\end{tabular}%
}
\caption{Mean Objective Intelligibility score (SIIB) and Confidence Interval (CI) for speech-shape noise at SNR -1 and SNR -3 for female and male target speaker. Higher is better. The best results are highlighted in bold. All pairs of systems are significantly different for $p\leq 0.005$, except the ones indicated by * and †. }
\label{tab:intelligililbity}
\end{table}

\section{Discussion}

The results show that \textit{Implicit modeling} via the Lombard style classifier was able to achieve similar results to models with explicit conditioning. This approach presents the advantage of not requiring domain knowledge and extensive linguistic and acoustics studies. We also observe that \textit{f0} extracted from male speakers in Lombard style has better impact on the intelligibility than from female speakers, which is consistent with the gender-specific changes in Lombard speech~\cite{junqua1993lombard}. On the other hand, our experiments on explicit modeling confirm previous studies on the importance of spectral tilt and energy for modeling the Lombard effect~\cite{ summers1988effects,junqua1993lombard}. We note that our models do not alter the durations, albeit observed in Lombard speech, which could further benefit the intelligibility. Additionally, this work is limited by the size of the dataset and the overall robotic prosody of the samples given the recording task, which also impacts the quality of the synthesized samples. Future work could explore the intelligibility enhancement on a larger and conversational recordings dataset. This in turn would allow us to train and evaluate TTS models trained on synthetic Lombard data generated with our model. We would also evaluate the intelligibility of systems on transcriptions and intelligibility scores targeted at hearing loss. Finally, adversarial learning could also be explored in future improvements to further disentangle source speaker identity from the explicit features, as well as grouping different source speakers by Lombard intensity to control the intelligibility level.

\section{Conclusion}
\label{sec:conclusion}

In this work, we analyze the impact of the Lombard effect on the Intelligibility of voices in noise and investigate Lombard-preserving voice conversion.  
Confirming previous studies, we show that the Lombard effect increases the intelligibility in noise, and that while naturalness is lost, speaker similarity can still be observed by listeners in noisy conditions. We investigate many-to-many voice conversion preserving Lombard style with both implicit and explicit conditioning. Spectral tilt and energy were the most beneficial features for the Lombard style, and we show that the model with the added reconstruction loss achieves intelligibility gain on par with the model conditioned on source features, while better preserving the speaker similarity.

\bibliographystyle{IEEE}
\bibliography{main}

\end{document}